# The Helmholtz Wave Potential:
# a non-probabilistic insight into Wave Mechanics

### Adriano Orefice[*], Raffaele Giovanelli, Domenico Ditto

*Università degli Studi di Milano - DISAA. Via Celoria, 2 - 20133 - Milan (Italy)*

### ABSTRACT

The behavior of classical monochromatic waves in stationary media is shown to be ruled by a novel, frequency-dependent function which we call *Wave* Potential, and which we show to be encoded in the structure of the Helmholtz equation. An exact, Hamiltonian, *ray-based* kinematical treatment, reducing to the usual eikonal approximation in the *absence* of Wave Potential, shows that its *presence* induces a mutual, perpendicular ray-coupling, which is *the one and only cause* of wave-like phenomena such as diffraction and interference. The Wave Potential, whose discovery does already constitute a striking novelty in the case of *classical* waves, turns out to play an even more important role in the *quantum* case. Recalling, indeed, that the time-independent Schrödinger equation (associating the motion of mono-energetic particles with stationary monochromatic *matter waves*) is itself a Helmholtz-like equation, the exact, ray-based treatment developed in the classical case is extended - *without resorting to statistical concepts* - to the exact, *trajectory-based* Hamiltonian dynamics of mono-energetic *point-like particles*. Exact, classical-looking particle trajectories may be defined, contrary to common belief, and turn out to be perpendicularly coupled by an *exact*, *energy-dependent* Wave Potential, similar in the form, but not in the physical meaning, to the *statistical, energy-independent* "Quantum Potential" of Bohm's theory, which is affected, as is well known, by the *practical necessity* of representing particles by means of *statistical wave packets*, moving along probability flux lines. This result, together with the connection shown to exist between Wave Potential and Uncertainty Principle, allows a novel, *non-probabilistic* interpretation of Wave Mechanics.

### RÉSUMÉ

On montre que le comportement des ondes monochromatiques classiques dans les milieux stationnaires est gouverné par une fonction dispersive, qu'on appelle ici "Potentiel d'Onde", codée dans la structure même de l'équation d' Helmholtz, et on présente une description exacte, en termes de rayons hamiltoniens (se réduisant à l'approximation de l'optique géométrique lorsque cette fonction est négligée), qui révèle une dépendance mutuelle entre les rayons et permet même le traitement de phénomènes typiquement ondulatoires tels que la diffraction et l'interférence. Puisque l'équation de Schrödinger indépendante du temps (associant le mouvement des particules mono-énergétiques avec des ondes materielles stationnaires monochromatiques) est elle-même une équation d'Helmholtz, le traitement mathématique valable dans le cas classique est étendu, sans recourir à des concepts probabilistes, à la dynamique hamiltonienne quantique de particules ponctiformes mono-énergétiques. Les trajectoires dynamiques exactes des particules sont couplées, dans ce cas, par un Potentiel d'Onde exact et dispersif similaire dans sa forme au "Potentiel Quantique" bohmien, dont il ne partage pas, cependant, la nature probabiliste et l'indépendance de l'énergie. Ce résultat, associé au lien entre le Potentiel d'Onde et le Principe d'Incertitude, suggère une nouvelle façon de comprendre la Mécanique Ondulatoire.

---

[*] Corresponding author -  adriano.orefice@unimi.it

## 1 - Introduction

*Let me say at the outset that I am opposing not a few special statements of quantum physics held today (1955): I am opposing the whole of it (...), I am opposing its basic views, shaped when Max Born put forward his **probabilistic interpretation**, which was accepted by almost everybody.* (E. Schrödinger, [**1**])

Let us say at the outset that the we put forward, in the present paper, a novel, **non-probabilistic interpretation** of Wave Mechanics.

The historical contingency which favored the "Copenhagen hegemony" in the **probabilistic interpretation** of Wave Mechanics was extensively reconstructed by Cushing, in 1994, in a very remarkable book [**2**]. The establishment of this hegemony after the publication of von Neumann's "impossibility" theorem [**3**], and the beginning of its end, are further, vividly expressed in Bell's words [**4**]: "*I relegated the  question to the back of my mind, and got on with more practical things, [until] in 1951 I saw the impossible done. It was in papers by David Bohm*" [**5**].

As reported in Refs. [**6,7**], however, Einstein himself (not to speak of the attitude of many other "founding fathers" of quantum physics) defined Bohm's theory "*too cheap*" for him, thus strongly contributing to its long-lasting refusal and to the Copenhagen hegemony - which did not prevent Bohm from maintaining and extending his standpoint, together with a few co-workers [**8-10**], until his death in 1992. After this date his theory survived in a somewhat esoteric line of thought [**11**], and finally emerged to a quite generally recognized official life during the last decade [**12-25**], mainly because of its applications to chemical physics and nano-technology.

The present work is based on the observation that Bohm's proposal of a "hidden variables" interpretation of Quantum Mechanics **was not radical enough**. While, indeed, in Bohm's words, "...*the use of a **statistical ensemble** is only (as in the case of classical statistical mechanics) a **practical necessity**, but not a reflection of an inherent limitation on the precision with which it is correct for us to conceive of the variables defining the state of the system*", **we avoid here such a practical necessity**, by means of a **non-statistical** approach where an *exact*, trajectory-based quantum dynamics, directly stemming from the Helmholtz-like structure of the time-independent Schrödinger equation, is developed without resorting to any kind of probabilistic concepts. We stick, in other words, *to the spirit, but not to the letter,* of Bohm's interpretation. No wave-packet, in particular, and no simultaneous solution of the time-dependent Schrödinger equation are required by our construction of the *exact* particle trajectories.

The basic key for this development is the transition (first performed in Ref.[**19**], and sketched in **Sect.2** of the present work) from the *approximate*, trajectory-based eikonal description of *classical* monochromatic waves to their *exact, and still trajectory-based*, Hamiltonian treatment. Both the *geometry* and the *kinematics* of the ray-trajectories (to which *no probabilistic concept* is attached) turn out to be mutually coupled by a **novel, exact, frequency-dependent "Wave**

**Potential**" function (encoded in the structure of the Helmholtz equation, and normally coupling the relevant, monochromatic ray trajectories) which is *the one and only cause* of typically wave-like phenomena such as diffraction and interference.

This mathematical treatment is extended, in **Sect.3**, from the *classical* to the *quantum* case, starting from the de Broglie-Schrödinger suggestion [**26-28**] that classical particle mechanics represents the *eikonal approximation* of a more general wave-like mechanics according to the basic Ansatz $\vec{p} = \hbar \, \vec{k}$, and from the consequent construction of the (Helmholtz-like) time-independent Schrödinger equation, associating the motion of *mono-energetic* particles with stationary *monochromatic* "matter waves". A **non-probabilistic**, trajectory-based Hamiltonian *quantum dynamics of point-like particles* is shown to be possible, under the rule of an exact, energy-dependent, trajectory-coupling **Wave Potential** function analogous to the one found in the classical case.

The relevant Hamiltonian equation system is numerically solved in **Sect.4**, in a *unified, dimensionless form* holding both for classical and quantum waves, in various cases of wave diffraction and interference, by means of a self-consistent approach where no simultaneous solution of the time-dependent Schrödinger equation is required.

The connection between the Wave Potential and the Uncertainty Principle is examined in **Sect.5**, showing both the origin and the limits of such a Principle.

Our *exact, stationary, energy-dependent* Wave Potential is shown in **Sect. 6** to have a physical meaning quite different from Bohm's *statistical, time-evolving, energy independent* Quantum Potential, in spite of their apparent formal coincidence. While indeed Bohm declares his belief, *in principle,* in the legitimacy of conceiving single particle trajectories [**5**], his description doesn't appear to differ so much, *in practice*, from the standard Copenhagen paradigm, with which it basically associates a set of probability flux lines representing weighted *statistical* averages (holding for travelling *wave-packets*) performed over the *exact*, mono-energetic trajectories of *point-like particles* provided by the present paper.

We conclude, in **Sects.7** and **8**, that the *time-independent* Schrödinger equation is not a simple particular case of the time-dependent one, but a quantum dynamical stage providing the "missing link" between classical particle dynamics and Bohm's theory. A link giving to Wave Mechanics a **novel, non-probabilistic** understanding, arising as a *direct and exact mathematical consequence* of the time-independent Schrödinger equation itself.

## 2 - The Helmholtz Wave Potential for classical waves

In the present Section, entirely devoted to *classical* waves, we assume both wave mono-chromaticity and stationary media, allowing the best theoretical and experimental analysis of diffraction and/or interference phenomena. Although our considerations may be easily extended to many kinds of classical (from acoustic to seismic) waves, we shall refer here, in order to fix ideas, to *classical electromagnetic waves* travelling through a stationary, isotropic and (generally) inhomogeneous dielectric medium according to a scalar wave equation of the simple form [**29**]

$$\nabla^2 \psi - \frac{n^2}{c^2} \frac{\partial^2 \psi}{\partial t^2} = 0 \quad , \tag{1}$$

where $\psi(x,y,z,t)$ represents any component of the electric and/or magnetic field, $n(x,y,z)$ is the (time independent) refractive index of the medium and $\nabla^2 \equiv \frac{\partial^2}{\partial x^2} + \frac{\partial^2}{\partial y^2} + \frac{\partial^2}{\partial z^2}$ . By assuming solutions of the form

$$\psi(x,y,z,t) = u(x,y,z)e^{-i\,\omega t} \tag{2}$$

we get the Helmholtz equation

$$\nabla^2 u + (n\,k_0)^2 u = 0 \, , \tag{3}$$

where $k_0 \equiv \frac{2\pi}{\lambda_0} = \frac{\omega}{c}$ . If we now perform the replacement

$$u(x,y,z,\omega) = R(x,y,z,\omega)\,e^{\,i\,\varphi\,(x,y,z,\omega)}, \tag{4}$$

with real $R$ and $\varphi$, and separate the real from the imaginary part, eq.(3) splits into the coupled system [**29**]

$$\begin{cases} (\vec{\nabla}\,\varphi)^2 - (n\,k_0)^2 = \dfrac{\nabla^2 R}{R} & \text{(a)} \\[2mm] \vec{\nabla} \cdot (R^2\,\vec{\nabla}\,\varphi) = 0 & \text{(b)} \end{cases} \tag{5}$$

where $\vec{\nabla} \equiv \partial/\partial\,\vec{r} \equiv (\partial/\partial x, \partial/\partial y, \partial/\partial z)$ and $\vec{r} \equiv (x,y,z)$ . The function $R(\vec{r},\omega)$ represents - with no intrinsically probabilistic meaning - the stationary wave amplitude distribution, and the second of eqs.(5) expresses the constancy of the flux of the vector $R^2\,\vec{\nabla}\,\varphi$ along any tube formed by the field lines of the wave vector

$$\vec{k} = \vec{\nabla}\,\varphi \, . \tag{6}$$

By introducing now the functions

$$D(\vec{r},\vec{k},\omega) \equiv \frac{c}{2\,k_0}[k^2 - (n\,k_0)^2 - \frac{\nabla^2 R}{R}] \tag{7}$$

and

$$W(\vec{r},\omega) = -\frac{c}{2k_0}\frac{\nabla^2 R(\vec{r},\omega)}{R(\vec{r},\omega)}, \qquad (8)$$

the differentiation of eq.(5a) leads to a Hamiltonian ray-tracing kinematic system of the form [**19, 25**]

$$\begin{cases} \dfrac{d\,\vec{r}}{d\,t} = \dfrac{\partial\,D}{\partial\,\vec{k}} = \dfrac{c\,\vec{k}}{k_0} \\[2mm] \dfrac{d\,\vec{k}}{d\,t} = -\dfrac{\partial\,D}{\partial\,\vec{r}} = \vec{\nabla}\,[\,\dfrac{c\,k_0}{2}\,n^2\,(\vec{r}) - W(\vec{r},\omega)] \end{cases} \qquad (9)$$

where $\partial/\partial\vec{k} \equiv (\partial/\partial k_x,\ \partial/\partial k_y,\ \partial/\partial k_z)$, and a ray velocity $\vec{v}_{ray} = \dfrac{c\,\vec{k}}{k_0}$ is implicitly defined. Notice that, as long as $k \equiv \left|\vec{k}\right| = k_0$, we'll have $v_{ray} \equiv \left|\vec{v}_{ray}\right| = c$.

The frequency-dependent function $W(\vec{r},\omega)$, which we call "*Helmholtz Wave Potential*", couples the ray trajectories of the monochromatic wave in a kind of self-refraction (which we call "*Helmholtz coupling*") affecting both their geometry and their kinematics. Such a function (which has the dimensions of a frequency) represents an intrinsic, dispersive property of the Helmholtz equation itself, and is determined by the amplitude space-distribution of the wave.

We observe, from eq.(5 b), that

$$\vec{\nabla}\bullet(R^2\,\vec{\nabla}\,\varphi) \equiv 2\,R\,\vec{\nabla}\,R\bullet\vec{\nabla}\varphi + R^2\,\vec{\nabla}\bullet\vec{\nabla}\varphi = 0. \qquad (10)$$

This equation has a double role:

- <u>On the one hand</u>, since no new trajectory may suddenly arise in the space region spanned by the considered wave trajectories, we must have $\vec{\nabla}\bullet\vec{\nabla}\,\varphi = 0$, so that $\vec{\nabla}\,R\bullet\vec{\nabla}\varphi = 0$: the amplitude $R(\vec{r},\omega)$ is distributed at any time (together with its functions and derivatives) on the wave-front reached at that time, and both $\vec{\nabla}\,R$ and $\vec{\nabla}\,W$ are perpendicular to $\vec{k} \equiv \vec{\nabla}\,\varphi$. The (frequency-dependent) coupling term $\vec{\nabla}\,W$ always acts, therefore, *perpendicularly* to the trajectories given by the ray-tracing system (9) for the assigned wave frequency. A basic consequence of this property is the fact that, in the case of electromagnetic waves propagating *in vacuo*, the absolute value of the ray velocity *remains equal to c all along each ray trajectory,* whatever its form may be, because such a term may only modify the direction, but not the amplitude, of the wave vector $\vec{k}$. For a general medium, the only possible changes of $k \equiv \left|\vec{k}\right|$ may be due to its refractive index *n(x,y,z)*, but not to the Wave Potential. The Helmholtz Wave Potential, in conclusion, has the

double property of being frequency-dependent (i.e. dispersive) and of coupling the relevant set of wave trajectories by acting, at each point, perpendicularly to them;

- <u>On the other hand</u>, thanks to the constancy of the flux of $R^2 \vec{\nabla} \varphi$, the function $R(\vec{r}, \omega)$, once assigned on the surface from which the wave is assumed to start, may be built up step by step, together with the Wave Potential $W(\vec{r}, \omega)$, along the ray trajectories. *The Hamiltonian kinematic system (9) is "closed", in other words, by eq.(5b),* so that the knowledge of the distribution of $R(\vec{r}, \omega)$ on a wave-front is the necessary and sufficient condition for the determination of its distribution on the next wave-front. This allows its numerical integration, and provides both an exact stationary system of coupled "rails" (which we shall call *Helmholtz trajectories*) orthogonal to the wave-fronts, and the ray motion laws along them, starting (with an assigned wave-vector) from definite points of the launching surface and coupled by the Wave Potential $W(\vec{r}, \omega)$.

In order to better specify this point, let us imagine a reference frame with a horizontal *z*-axis directed towards the right hand side, where a plane monochromatic wave with an amplitude distribution $R(x, y, \omega) = const$ is incoming from the left. According to eq.(8), the Wave Potential will remain equal to zero, all over the region spanned by the wave, as long as $R$ remains a constant. The presence of an inhomogeneous refractive index $n(\vec{r})$, however, shall be sufficient, in general, to cause an inhomogeneous amplitude distribution $R(\vec{r}, \omega)$, and therefore a non-vanishing Wave Potential. Let us also imagine that the wave encounters an optical system of some kind. Although a Michelson interferometer, or even a single slit on a vertical screen, would be adequate examples in order to fix ideas, we have rather in mind a more general arrangement consisting of lenses, mirrors, diaphragms, slits and so on, allowing to determine the wave amplitude profile $R(\vec{r}, \omega)$ over an arbitrary starting front - such as, for instance, a vertical plane, whence the wave is launched toward the right, with $k_z = k_0 > 0$. The knowledge of this launching profile supplies eqs.(9) and (5b) with the necessary and sufficient information for the construction of the successive wave-fronts. Since, moreover, the absolute value of the wave intensity is inessential for this construction, we find it appropriate to speak not so much of a "wave beam", as of a virtual stationary set of Helmholtz trajectories, i.e. of a set of fixed rails, normal to the wave-fronts, along which the wave rays are channeled. *Unusual, indeed, as it could seem at first sight*, one has basically in mind (when speaking, for instance, of a Gaussian wave beam) a well defined virtual set of stationary Helmholtz "rails", coupled by the Helmholtz Wave Potential, along which waves of any possible (even vanishing) intensity may be launched.

When, in particular, the space variation length $L$ of the wave amplitude $R(\vec{r}, \omega)$ turns out to satisfy the condition $k_0 L >> 1$, eq.(5a) is well  approximated by the "eikonal equation" [**29**]

$$(\vec{\nabla} \varphi)^2 \simeq (n k_0)^2, \tag{11}$$

decoupled from eq.(5 b), and the term containing the Wave Potential may be dropped from the ray-tracing system (9). In this *eikonal* (or "geometrical optics") approximation the rays, both in the form of their trajectories and in their kinematics, are no longer mutually coupled by the Wave Potential, and propagate independently from one another under the only influence of the external refractive index. The main consequence of this independence is the absence, in such a limiting case, of typically wave-like phenomena such as diffraction and/or interference, which are only due to a non-vanishing Wave Potential.

Let us remind here that while our equations (9) provide an *exact* Hamiltonian description of the wave kinematics, an *approximate* Hamiltonian treatment was presented in 1993/94 by one of the Authors (A.O., [**30**,**31**]), for the quasi-optical propagation of electromagnetic Gaussian beams at the electron-cyclotron frequency in the magnetized plasmas of Tokamaks such as JET and FTU, and applied in recent years by an *équipe* working on the Doppler back-scattering microwave diagnostics installed on the Tokamak TORE SUPRA of Cadarache [**32**]. A complex eikonal equation, amounting to a first order approximation of the beam diffraction, was adopted in Refs.[**30-32**] in order to overcome the collapse, for narrow microwave beams, of the ordinary geometrical optics approximation.

Coming back, now, to the most general case, let us finally observe that if we pass to *dimensionless variables* by expressing:

- the space variable $\vec{r}$ (together with the space operators $\vec{\nabla}$ and $\nabla^2$) in terms of an "*a priori*" arbitrary physical length $w_0$ , to be specified later on,
- the wave vector $\vec{k}$ in terms of $k_0$, and
- the time variable $t$ in terms of $w_0/c$ ,

and maintaining for simplicity the names $\vec{r}, \vec{k}, t,$ for the new variables, the Hamiltonian system (9) takes on the dimensionless form

$$\begin{cases} \dfrac{d\,\vec{r}}{d\,t} = \vec{k} \\ \dfrac{d\,\vec{k}}{d\,t} = \dfrac{1}{2}\vec{\nabla}\,[\,n^2 + \left(\dfrac{\varepsilon}{2\pi}\right)^2\,G(x,y,z)] \end{cases} \tag{12}$$

where

$$\varepsilon = \lambda_0 \big/ w_0 \tag{13}$$

and the Wave Potential (with opposite sign) is represented by the (dimensionless) function

$$G(x,y,z) = \dfrac{\nabla^2 R}{R}, \tag{14}$$

where it must be borne in mind that the $\vec{\nabla}$ operator is expressed in dimensionless terms.

Notice that different values of $\varepsilon \equiv \lambda_0 / w_0$ (i.e. *different frequencies* $\omega = 2\pi c / \lambda_0$, for a fixed value of the assumed unit of length, $w_0$) lead to different values of the coefficient weighting the effect of the potential function G, and therefore to *different trajectories*. In this sense we may speak of a *dispersive* character of the *Wave Potential* itself. For a fixed value of $w_0$, waves of different frequencies travel along different trajectories, to which the coupling action of the Wave Potential maintains itself perpendicular.

## 3 - The Helmholtz Wave Potential in Wave Mechanics

Let us pass now to the case of mono-energetic, non-interacting particles of mass $m$ launched with an initial momentum $\vec{p}_0$ into a force field deriving from a stationary potential energy $V(\vec{r})$. The *classical* motion of each particle may be described, as is well known [**29**], by the time-independent Hamilton-Jacobi equation

$$(\vec{\nabla} S)^2 = 2 \, m \, [E - V(\vec{r})] \quad , \tag{15}$$

where $E = p_0^2 / 2m$ is the total energy of the particle, and the basic property of the function $S(\vec{r})$ is that the particle momentum is given by

$$\vec{p} = \vec{\nabla} S . \tag{16}$$

The analogy between eqs.(15) and (11), and between eqs. (16) and (6), suggested to de Broglie [**26**] and Schrödinger [**27, 28**], as is very well known, that classical particle dynamics could represent the geometrical optics approximation of a more general wave-like reality, described by a suitable *Helmholtz-like* equation in terms of monochromatic *matter waves* associated to monochromatic particles. Recalling, indeed, de Broglie's fundamental Ansatz, $\vec{p} = \hbar \vec{k}$, such an equation is immediately obtained [**33**] from eq.(3) by means of the replacements

$$\varphi \to \frac{S}{\hbar} \; ; \quad n^2(\vec{r}) \to 1 - \frac{V(\vec{r})}{E} \; ; \quad k_0 \equiv 2\pi / \lambda_0 \to p_0 / \hbar ,$$

transforming it into the standard *time-independent* Schrödinger equation for a stationary potential field $V(\vec{r})$:

$$\nabla^2 u + \frac{2m}{\hbar^2} [E - V(\vec{r})] \, u = 0 \quad . \tag{17}$$

By applying to eq.(17) the same procedure leading from the Helmholtz eq.(3) to eqs.(5), and assuming

$$u(\vec{r},E) = R(\vec{r},E)\, e^{\, i\, S(\vec{r},E)\, /\, \hbar}, \qquad (18)$$

eq.(17) splits into the well-known [**34**] coupled system

$$
\begin{cases}
(\vec{\nabla}\, S)^2 - 2m(E - V) = \hbar^2\, \dfrac{\nabla^2 R}{R} & (\text{a})\\[2mm]
\vec{\nabla}\cdot(R^2\, \vec{\nabla}\, S) = 0 & (\text{b})
\end{cases}
\qquad (19)
$$

strictly analogous to the system (5). After having introduced the functions

$$H\,(\vec{r},\,\vec{p},E\,) = \frac{p^2}{2m} +\, V(\vec{r}) + Q(\vec{r},E) \qquad (20)$$

and

$$Q(\vec{r},E) = -\,\frac{\hbar^2}{2m}\,\frac{\nabla^2 R(\vec{r},E)}{R(\vec{r},E)} \qquad (21)$$

we obtain now [**19, 25**], by differentiating eq.(19a), the Hamiltonian *quantum dynamical* system, strictly analogous to the *classical kinematic* system (9),

$$
\begin{cases}
\dfrac{d\,\vec{r}}{d\,t} = \dfrac{\partial\,H}{\partial\,\vec{p}} = \dfrac{\vec{p}}{m}\\[3mm]
\dfrac{d\,\vec{p}}{d\,t} = -\dfrac{\partial\,H}{\partial\,\vec{r}} = -\vec{\nabla}[V(\vec{r}) + Q(\vec{r},E)]
\end{cases}
\qquad (22)
$$

where $\partial/\partial\,\vec{p} \equiv (\partial/\partial\,p_x,\,\partial/\partial\,p_y,\,\partial/\partial\,p_z)$. The function $Q(\vec{r},E)$, just like $V(\vec{r})$, is a stationary function of position [**35**]. It's seen to have the same basic structure and mathematical role of the Wave Potential function $W(\vec{r},\omega)$, and to be *formally* coincident, moreover, with the well known Quantum Potential of Bohm's theory [5]: *there exist, however, fundamental differences* between the energy-dependent function $Q(\vec{r},E)$ and Bohm's energy-independent Quantum Potential $Q_B(\vec{r},t)$, which shall be stressed in Section 6.

Because of its analogy with the function $W(\vec{r},\omega)$, we shall refer to the function $Q(\vec{r},E)$, for brevity sake, *with the same name of "Wave Potential"*. It doesn't represent so much, indeed, a "quantum" as a "wave" feature, due to the wave properties of matter. It must be observed, however, that $W(\vec{r},\omega)$ has the dimensions of a *frequency*, while $Q(\vec{r},E)$ has the dimensions of an *energy*.

Because of their strict analogy, we shall submit the *quantum* dynamical Hamiltonian system (22) to the same interpretation and mathematical treatment applied in the case of the system (9), holding for *classical* electromagnetic waves.

Let us observe that de Broglie's Ansatz $\vec{p} = \hbar \vec{k}$ replaces, *and frees from any probabilistic implication*, what in the de Broglie-Bohm theory **[5, 36]** is called "the guidance equation assumption".

Once more, the function $R(\vec{r}, E)$ provides the amplitude distribution, over the wave front, of a mono-energetic wave, with no intrinsically probabilistic meaning. The presence of the potential $Q(\vec{r}, E)$ causes, in its turn, the "*Helmholtz coupling*" of the quantum wave trajectories, and its absence or omission would reduce the quantum equation system (22) to the standard classical set of dynamical equations, which constitute therefore, as expected **[26-28]**, its geometrical optics approximation. We find it very important to stress here, indeed, that while a *classical limit* is often claimed to correspond to the (scarcely significant) limit $\hbar \rightarrow 0$ **[33, 34]**, the *classical limit* is clearly seen to be approached, instead, when the role of the Wave Potential may be neglected, i.e. - if $L$ is the space variation length of $R(\vec{r}, E)$ - in the eikonal limit $k_0 L \gg 1$.

In complete analogy with the classical electromagnetic case of the previous Section,

1) the (energy-dependent) wave-mechanical "force" $-\vec{\nabla} Q(\vec{r}, E)$ turns out to act *perpendicularly* to the trajectories of the relevant particles. Because of its perpendicularity to $\vec{p} \equiv \vec{\nabla} S$ it cannot modify the amplitude of the particle momentum (while modifying, in general, its direction), so that *no energy exchange* may ever occur between particles and Wave Potential;

2) the relation (19 b) allows to obtain both $R(\vec{r}, E)$ and $Q(\vec{r}, E)$ along the particle trajectories, *thus providing the "closure" of the dynamical Hamiltonian system (22)* and making its integration possible without resorting to the simultaneous solution of a time-dependent Schrödinger equation. While in Bohm' theory " *the need for a probability description is not regarded as inherent in the very structure of matter*" [5], but the use of *statistical ensembles*, based on the time-dependent Schrödinger equation, is viewed *as a practical necessity,* no *statistical* concept, and no wave-packet, are employed in our treatment, where the knowledge of the distribution of $R(\vec{r}, E)$ on a wave-front is the necessary and sufficient condition, thanks to eq.(19b), to determine the distribution over the next wave-front and to provide both an *exact* stationary system of coupled "rails" and the particle dynamical laws along them. In complete analogy with the case of electromagnetic waves treated in Sect.2, any experimental set-up would be associated with a specific set of wave-fronts, and therefore with a stationary set of coupled trajectories (orthogonal to these wave-fronts), along which particles are channeled according to the dynamical system (22).

Notice, in this respect, that stationary particle trajectories (orthogonal to the surfaces $S(\vec{r}) = const$) are already present in the corresponding *classical* Hamilton dynamics: the only, but fundamental, difference, in the *wave-mechanical* case, is represented by the (Helmholtz) trajectory coupling.

3) Each particle is endowed, at any time, with a well defined momentum, associated with its instantaneous (point-like) position. The equation system (22) describes, in other words, the exact dynamics of mono-energetic, classical-looking *point-like particles* starting from assigned point-like positions, and following a well defined set of trajectories, coupled by $Q(\vec{r}, E)$. Let us remind here that while our equations (22) provide an *exact and general* Hamiltonian description of the *wave-mechanical* particle trajectories, an *approximate* Hamiltonian treatment was presented in 1997 by one of the Authors (A.O., [**37**]), for the particular case of *Gaussian* particle beams. A complex eikonal equation, amounting to a first order approximation of the beam diffraction, was adopted there, in strict analogy with the one employed in the *classical* electromagnetic case of Refs.[**30,31**], in order to overcome the collapse, for narrow beams, of the ordinary geometrical optics approximation. In complete analogy with the previous, *classical* electromagnetic case, the *wave-mechanical* Hamiltonian system (22) may be put in a suggestive *dimensionless* form by expressing lengths (as well as $\vec{\nabla}$ and $\nabla^2$) in terms of a physical length $w_0$ ( to be defined later on), momentum in terms of $p_0$ and time in terms of $w_0 / v_0$, with $v_0 = p_0 / m$:

$$
\begin{cases}
\dfrac{d\,\vec{r}}{d\,t} = \vec{p} \\[2mm]
\dfrac{d\,\vec{p}}{d\,t} = \dfrac{1}{2} \vec{\nabla} \left[ -\dfrac{V}{E} + (\dfrac{\varepsilon}{2\pi})^2 \; G(x,y,z) \right]
\end{cases}
\tag{23}
$$

where the parameter $\varepsilon$ and the (dimensionless) potential function *G(x,y,z)* are given, once more, by eqs.(13) and (14), respectively. Not surprisingly, the *quantum* dimensionless system (23) turns out to formally coïncide with the *classical* dimensionless system (12), by simply replacing $\vec{k}$ by $\vec{p}$ and $n^2$ by *(1-V/E)*. The trajectory coupling due to *G(x,y,z)* is therefore a physical phenomenon affecting *both classical and quantum waves*, and its absence would reduce the relevant equations to the ones, respectively, of standard geometrical optics and of classical dynamics. Let us observe once more that different values of the parameter $\varepsilon \equiv \lambda_0 / w_0$, i.e. (for a fixed value of the assumed unit of length, $w_0$), different values of the total energy $E = \dfrac{2}{m}(\pi \hbar / \lambda_0)^2$, lead to different sets of trajectories, i.e. to a dispersive behavior.

## 4 - Unified numerical computations

Once assigned on the launching surface of the wave, the wave amplitude profile *R(x,y,z)* and the consequent potential function $G(x,y,z)$ may be numerically built up step by step, together with their derivatives, along the wave trajectories. We present here some applications of the *dimensionless* Hamiltonian systems (12) and/or (23) to the propagation of collimated beams injected at $z = 0$, parallel to the $z$ - axis, simulating wave *diffraction* and/or *interference* through suitable slits, each one of half width $w_0$. *Here we perform, therefore, the choice of the length* $w_0$, and we assume $\varepsilon \equiv \lambda_0 / w_0 < 1$.

The problem is faced by taking into account, for simplicity sake, either (*quantum*) particle beams in the absence of external fields ($V = 0$) or (*classical*) electromagnetic beams *in vacuo* ($n^2 = 1$), with a geometry allowing to limit the computation to the $(x,z)$-plane. Because of the coincidence between the (dimensionless) Hamiltonian systems (12) and (23), the only choice to be performed is *between the names,* $\vec{k}$ or $\vec{p}$, of the variables - and we opt here for the second one, reminding however that we are not necessarily speaking of quantum topics.

Recalling that, because of the transverse nature of the gradient $\vec{\nabla} G$, the *amplitude* of $\vec{p}$ remains unchanged (in the absence of external fields and/or refractive effects) along each trajectory, we have

$$\begin{cases} p_x(t = 0) = 0; \ \ p_z(t = 0) = 1 \\ p_z(\,t \geq 0) = \sqrt{1 - p_x^{\,2}(\,t \geq 0)} \end{cases} \qquad (24)$$

and the dimensionless Hamiltonian system (23) reduces to the form

$$\begin{cases} \dfrac{d\,x}{d\,t} = \ p_x \\[2mm] \dfrac{d\,z}{d\,t} = \ \sqrt{1 - p_x^{\,2}} \\[2mm] \dfrac{d\,p_x}{d\,t} = \ \dfrac{\varepsilon^{\,2}}{8\,\pi^{\,2}} \ \dfrac{\partial\,G(x,z)}{\partial\,x} \end{cases} \qquad (25)$$

where (in dimensionless terms)

$$G(x,z) \equiv \frac{\nabla^2 R}{R} = \frac{\partial^2 R\,/\,\partial\,x^2}{p_z^{\,2}\ R} \qquad , \qquad (26)$$

and the Hamiltonian system (25) is "closed", as usual,  by eq.(19 b). Considering, in the present case, a set of  rays labelled by the index  $j$, if:

1) $x_j(t)$ and $z_j(t)$  are the coordinates of the point reached by the *j-th* ray at the time $t$;  2) $R_j(t)$ is the value assumed by the wave amplitude along such a  ray at the time $t$; and

3) $\qquad\qquad\qquad d_j(t) = \sqrt{(x_j(t) - x_{j-1}(t))^2 + (z_j(t) - z_{j-1}(t))^2} \qquad (27)$

is the distance between two adjacent rays at the same time step,

the "closure" equation may be written in the form

$$R_j^2(t)\,d_j(t) = const\,. \tag{28}$$

We assumed throughout the numerical computations the value $\varepsilon \equiv \lambda_0 / w_0 = 1.65 \times 10^{-4}$. Let us mention, for comparison, that a case of cold neutron diffraction was considered in **Ref.[12]** with

$$\lambda_0 = 19.26 \times 10^{-4}\,\mu m\,,\quad 2w_0 = 23\mu m\,,$$
$$\varepsilon = \lambda_0 / w_0 \cong 1.67 \times 10^{-4} \tag{29}$$

The wave launching amplitude distribution $R(x;z=0)$ (from whose *normalization* the function $G$ is obviously independent) was assigned, in Refs.[**19**, **25**, **38**], by means of suitable superpositions of Gaussian functions allowing a wide variety of beam profiles and an arbitrary number of "slits".

The values of $R(x;z>0)$ were then computed step by step by means of a symplectic integration method, and connected, at each step, by a Lagrange interpolation, allowing to perform space derivatives and providing both $G(x;z>0)$ and the full set of trajectories.

Figs.1 and 2 present, respectively, the initial (continuous) and final (dashed) transverse wave intensity profiles, and the set of particle trajectories, in the case of a *(non-Gaussian)* wave beam launched along the *z*-axis and diffracted by a single slit of half-width $w_0$, centered at $x=0$ on a vertical plane screen at $z=0$.

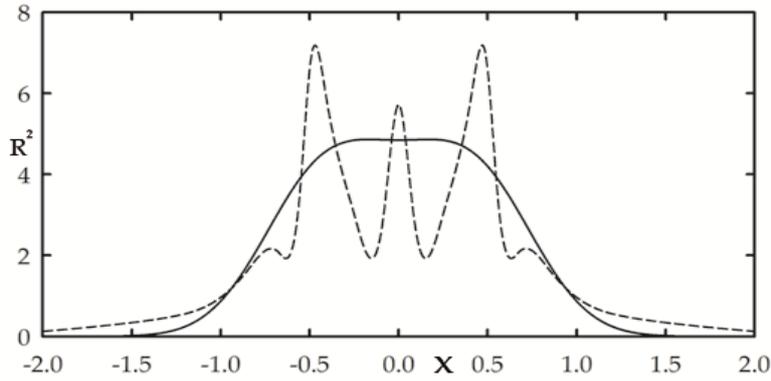

**Fig.1** - Initial (continuous) and final (dashed) transverse profiles of the beam intensity (in arbitrary units) for a general diffraction case.

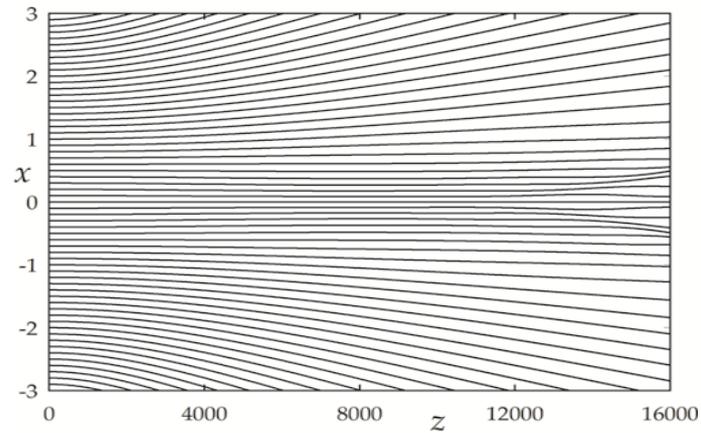

**Fig 2** - Particle trajectories on the $(x,z)$-plane relevant to Fig.1.

Fig.3 and 4 present, in their turn, the initial (continuous) and final (dashed) transverse beam intensity profiles, and the relevant set of beam trajectories on the $(x,z)$-plane, for an interference two-slits case.

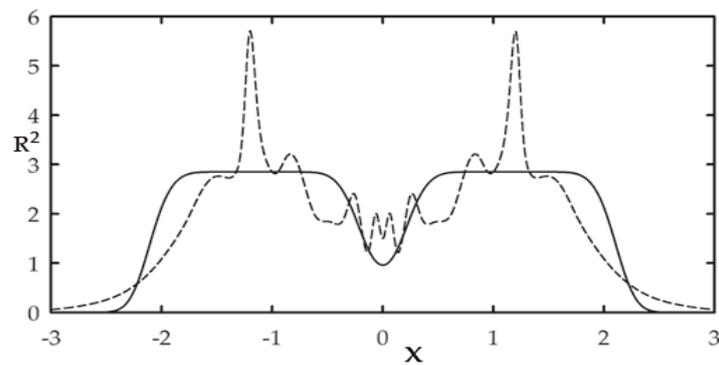

**Fig.3** Initial (continuous) and final (dashed) transverse profiles of the beam intensity for a two-slits interference case.

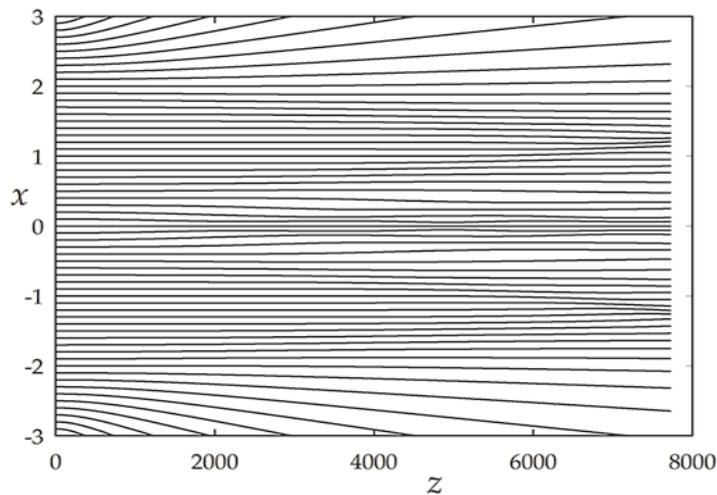

**Fig.4** - Wave trajectories relevant to Fig.3.

### 5 - Quantum Uncertainty, revisited

Warning that we shall refer, *within the present Section*, to **dimensioned** space and momentum variables, let us consider now the diffraction case of a **Gaussian quantum particle beam**, launched  along the *z*-axis (from a slit centered at  *x = 0* and placed on a vertical screen at *z = 0*) with initial momentum components

$$p_x(t=0)=0; \ p_z(t=0)=p_0=\hbar k_0=2\pi\hbar/\lambda_0 \qquad (30)$$

and with an initial space amplitude *(x,z)*-distribution of the form

$$R(x;z=0) \div exp(-x^2/w_0{}^2). \qquad (31)$$

The length  $w_0$ , representing the distance from the symmetry axis at which the beam amplitude is reduced by a factor *1/e*, is conventionally called "*half-width*" of the beam. The trajectory pattern obtained by the numerical integration of the quantum dynamical system (25), is shown in Fig.5, where the two heavy lines represent the so-called *waist-lines* of the Gaussian beam, given by the relation [**39**]

$$x(z)=\pm\sqrt{w_0{}^2+\left(\frac{\lambda_0\,z}{\pi\,w_0}\right)^2} \qquad , \qquad (32)$$

concerning the quasi-optical *paraxial* approximation, and representing in the present case the trajectories starting (at $z = 0$) from the positions $x = \pm \ w_0$. Notice that the excellent agreement between the *analytical* expression (32) and our *numerical* results provides a crucial test of our approach and interpretation.

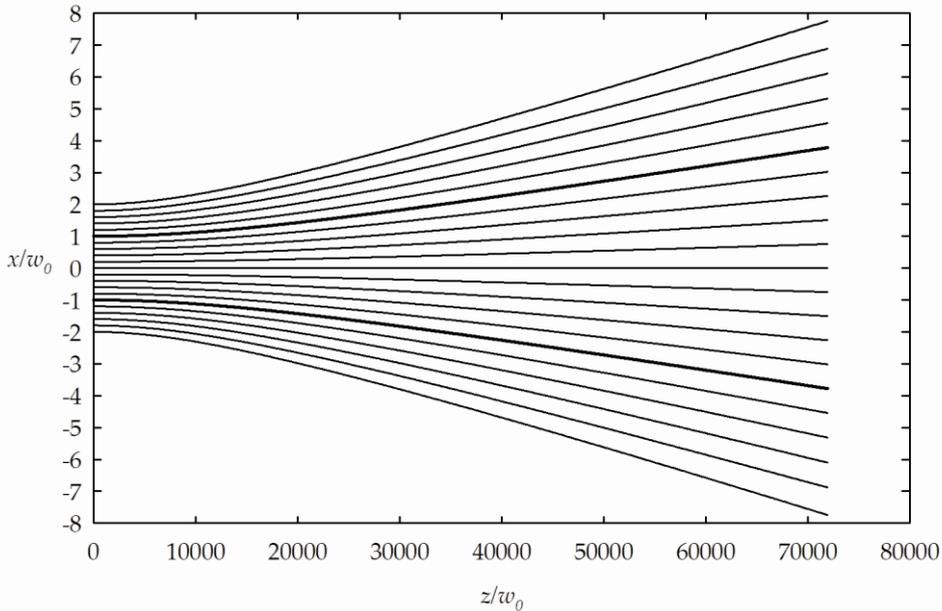

**Fig.5** – Diffracted trajectories, and waist lines, of a **Gaussian beam** on its vertical symmetry plane.

The passage of a particle through a narrow slit is a well known procedure for its *space* localization, providing a *space uncertainty* $\Delta x \approx 2 w_0$. After having crossed the slit, the beam maintains an almost collimated structure, with $p_x \approx 0$, as long as $z \ll \pi w_0^2 / \lambda_0$, diverging then, for $z \gg \pi w_0^2 / \lambda_0$, between the symmetric limiting slopes $\dfrac{x}{z} \approx \pm \dfrac{\lambda_0}{\pi w_0}$. A transverse $p_x$ component, ranging between $p_x \approx \pm 2 \hbar / w_0$, is therefore *progressively* developed, under the cumulative action of the Wave Potential, with a *momentum uncertainty* $\Delta p_x \approx 4 \hbar / w_0$, leading to the suggestive *asymptotic* relation

$$\Delta x \, \Delta p_x \approx 8 \hbar > h \tag{33}$$

Clearly enough, this relation turns out to be *violated* for $z < \pi w_0^2 / \lambda_0$, i.e. close enough to the slit location, where $\Delta p_x \approx 0$. The Uncertainty Relation *doesn't appear, therefore, to be a general and intrinsic property of physical reality*, but a local and limited effect - just like diffraction and interference - of the Wave Potential, due to our experimental information.

To be sure: any possible *space uncertainty* may only be due to our (not unavoidable) ignorance of the starting (point-like) positions of particles, and the consequent *momentum uncertainty* is only an *asymptotic* ($z \gg \pi w_0^2 / \lambda_0$) effect of the Wave Potential: an effect which turns out to be negligible in the space region $z < \pi w_0^2 / \lambda_0$.

## 6 - Bohm's Quantum Potential

Coming now to a comparison with Bohm's approach, let us previously recall that, starting from eqs.(2) and (17), one gets

$$\nabla^2 \psi - \frac{2m}{\hbar^2} V(\vec{r}) \, \psi = -\frac{2m}{\hbar^2} E \, \psi \equiv -\frac{2mi}{\hbar} \frac{E}{\hbar \omega} \frac{\partial \psi}{\partial t}, \tag{34}$$

an equation which, by assuming the Planck relation

$$E = \hbar \omega, \tag{35}$$

- i.e. by attributing to the total energy of a material particle a property (not needed in the *time independent* case (17))  holding, *stricto sensu*, in radiation theory - reduces to the usual form of the *time-dependent* Schrödinger equation for a stationary potential field $V(\vec{r})$,

$$\nabla^2 \psi - \frac{2m}{\hbar^2} V(\vec{r}) \psi = -\frac{2m\,i}{\hbar}\ \frac{\partial \psi}{\partial t}, \tag{36}$$

where $E$ and $\omega$ are not explicitly involved, and no wave dispersion is therefore, in principle, described. As is well known [**33**, **34**], indeed, the diffusion-like eq.(36) (representing a rare example of an intrinsically complex equation in physics) is not a wave equation: its wave-like implications are only due to its connection with eq.(17). We recall however that, just like the Helmholtz equation (3) is associated with the wave equation (1), the Helmholtz-like eq.(17) is associated - *via* eqs.(2) and (35) - with the ordinary-looking wave equation [**33**]

$$\nabla^2 \psi = \frac{2m}{E^2} (E-V) \frac{\partial^2 \psi}{\partial t^2} \tag{37}$$

providing significant information about the wave propagation and the dispersive character of a mono-energetic matter wave. To be sure, the most significant information is the fact itself that mono-energetic (i.e. monochromatic) *matter waves* propagate exactly like all other monochromatic waves.

While - if we accept eq.(35) - eq.(36) is a mathematical truism, its "stronger" version:

$$\nabla^2 \psi - \frac{2m}{\hbar^2} V(\vec{r},t) \psi = -\frac{2m\,i}{\hbar}\ \frac{\partial \psi}{\partial t}, \tag{38}$$

containing a time dependent potential, $V(\vec{r},t)$, *may only be assumed as a separate Ansatz*. We could even say that - because of the absence of stationary eigen-states - eq.(38) does not directly describe wave-like features: quite a *paradox* for what is assumed to be the basic equation of Wave Mechanics. Its main justification is given by various plausibility arguments (see, for instance, Ref.[**40**]) and by its current application in fields such as molecular [**41**] and ultra-fast laser [**42**, **43**] dynamics. Bohm's approach [**5**] performs, as is well known, a replacement of the form

$$\psi(\vec{r},t) = R(\vec{r},t)\, e^{\,i\,S(\vec{r},t)\,/\,\hbar} \tag{39}$$

into eq.(38) itself, splitting it, after separation of real and imaginary parts, into the equation system

$$\begin{cases} \dfrac{\partial P}{\partial t} + \vec{\nabla} \cdot \left(P\,\dfrac{\vec{\nabla} S}{m}\right) = 0 \\[2mm] \dfrac{\partial S}{\partial t} + \dfrac{(\vec{\nabla} S)^2}{2m} + V(\vec{r},t) - \dfrac{\hbar^2}{2m} \dfrac{\nabla^2 R}{R} = 0 \end{cases} \tag{40}$$

where, in agreement with the standard Copenhagen interpretation, the function $P \equiv R^2$ is assumed to represent, in Bohm's own terms, the *probability density for particles belonging to a statistical ensemble.*

The first of eqs.(40) is viewed, in Bohm's theory, as a fluid-like continuity equation for such a *probability density*, while the second one - having the form of *a dynamic* Hamilton-Jacobi equation, but including a *statistical, time-evolving, energy independent* "Quantum Potential"

$$Q_B(\vec{r},t) = -\frac{\hbar^2}{2m} \frac{\nabla^2 R(\vec{r},t)}{R(\vec{r},t)},$$ (41)

to be compared with the *exact, stationary, energy dependent* Wave Potential $Q(\vec{r},E) = -\frac{\hbar^2}{2m} \frac{\nabla^2 R(\vec{r},E)}{R(\vec{r},E)}$ of eq.(21) - is viewed as suggesting that "*precisely definable and continuously varying values of position and momentum*" [**5**] may be associated, ***in principle***, with each particle. Since however, according to Bohm, "*the most convenient way of obtaining R and S is to solve for the Schrödinger wave function*" [**5**], we are led, ***de facto***, to an *unavoidably statistical description* of the particle motion.

The situation is even more evident if we limit our attention to a *stationary* external potential $V(\vec{r})$. In this case the time-independent Schrödinger equation (17) admits in general, as is well known [**33**, **34**], a (discrete or continuous, according to the boundary conditions) set of energy eigen-values and orthonormal eigen-modes which (referring for simplicity to the discrete case) we shall call, respectively, $E_n$ and $u(\vec{r},E_n) \equiv u_n(\vec{r})$.

If we make use of eqs.(2) and (35), and define the eigen-frequencies $\omega_n \equiv E_n / \hbar$, together with the eigen-functions

$$\psi_n(\vec{r},t) = u_n(\vec{r}) e^{-i\omega_n t} \equiv u_n(\vec{r}) e^{-i\frac{E_n}{\hbar}t}$$ (42)

and with an arbitrary linear superposition of them,

$$\psi(\vec{r},t) = \sum_n c_n \psi_n$$ (43)

(with constant coefficients $c_n$), we may verify that such a superposition, when inserted into the *time-dependent* Schrödinger equation (36), reduces it to the form

$$\sum_n c_n e^{-i\frac{E_n}{\hbar}t} \left\{ \nabla^2 u_n + \frac{2m}{\hbar^2} [E_n - V(\vec{r})] u_n \right\} = 0$$ (44)

showing, by comparison with eq.(17), that eq.(43) provides a general solution of equation (36) itself. Written in duly normalized form, this solution was given by Born [**44**] an *ontologically probabilistic interpretation* which, even though "*no generally accepted derivation has been given to date*" [**45**], has become one of the standard principles of Quantum Mechanics.

Eq.(43) represents however, in any case, a weighted average over the full set of $\psi_n$, where the relative weights $c_n$ are determined by the available physical information.

It is worth noticing that while the eigen-functions $\psi_n$ may be righteously called "wave functions", such a name would be somewhat misleading for the function $\psi(\vec{r}, t)$ defined in eq.(43) and solving eq.(36), which, as is well known, is not even a wave equation.

The construction, now, of Bohm's flux lines and motion laws requires once more the simultaneous solution of the time-dependent Schrödinger equation, describing, in the present case, the diffusive evolution of a physical information assigned in the statistical form of a wave packet, i.e. as a weighted average, evolving in time. Although the use of *statistical ensembles* is claimed to be viewed [**5**] "*as a practical necessity, and not as a manifestation of an inherent lack of determination of the particle nature and motion*", any feature due to monochromatic properties such as dispersion and transverse trajectory coupling is *de facto* absent, or smoothed, in Bohm's treatment, because of its average character: the sum (43) over the full set of mono-energetic eigen-functions hinders, in fact, the possibility of distinguishing their individual peculiarities, so that the description, for instance, of diffraction and interference features - which is immediate in the monochromatic case -would require, for a wave packet, a very careful choice of the set of parameters $c_n$, strictly centering $R(\vec{r}, t)$ around a particular $\psi_n$.

Bohm's approach, in conclusion, doesn't appear to differ so much from the standard Copenhagen paradigm, with which it associates a set of fluid-like probability flux lines representing an average over a set of exact mono-energetic trajectories, *conjectured* by Bohm and *determined* in the present paper.

Let us notice that in the stationary form of the Hamilton-Jacobi dynamical equation we have

$$\frac{\partial S}{\partial t} = -E \,, \tag{45}$$

so that Bohm's equations (40) reduce to the system

$$\begin{cases} \vec{\nabla} \cdot (P \dfrac{\vec{\nabla} S}{m}) = 0 \\ -E + \dfrac{(\vec{\nabla} S)^2}{2m} + V(\vec{r}) - \dfrac{\hbar^2}{2m} \dfrac{\nabla^2 R}{R} = 0 \end{cases} \tag{46}$$

*seemingly* endowed with the same physical content of our eqs.(19), and *seemingly* suggesting, therefore, that eqs.(19) are only a particular and minor case of Bohm's *probabilistic* equations (40). *This, however, is not the case*. Eqs.(19), in fact, are a direct consequence of the (Helmholtz-like) time-independent Schrödinger equation (17), just like eqs.(5) are a direct consequence of the Helmholtz eq.(3), with which no probabilistic concept is associated.

The time-independent Schrödinger equation provides therefore, by means of the Hamiltonian system (22), stemming from eqs.(19), the *exact*, *non-probabilistic* quantum dynamics which Bohm's treatment *could not* achieve.

## 7- Summary and discussion

With an apparent delay of one century with respect to the development of standard Analytical Mechanics, the Authors of the present paper showed, *for the first time*, that an *exact, ray-based* Hamiltonian description of *classical* monochromatic waves - holding, *contrary to a wide-spread opinion*, even for typically wave-like phenomena such as diffraction and interference - may be obtained in terms of a suitable "Wave Potential" function, encoded in the structure of the Helmholtz equation. The main role of such a function is to cause a mutual ray-coupling, perpendicular to the rays themselves, which is the one and only cause of any deviation from geometrical optics.

Recalling, then, that the time-independent Schrödinger equation (associating the motion of mono-energetic particles with stationary monochromatic "matter waves") is itself a Helmholtz-like equation, this treatment was extended to develop an *exact*, *trajectory-based*, Hamiltonian *quantum dynamics* of *point-like* particles, reducing to usual *classical dynamics* in the absence of a Wave Potential. Keeping in mind that, in Bohm's words, *"...the use of a statistical ensemble is (as in the case of classical statistical mechanics) only a practical necessity, and not a reflection of an inherent limitation on the precision with which it is correct for us to conceive of the variables defining the state of the system"* [**5**], we stress that while our approach shares Bohm's *philosophical* position, it overcomes that *practical necessity* by means of an *exact*, trajectory-based quantum dynamics, avoiding any use of statistical wave packets.

## 8 - Conclusions

We mention here a reflection due to E.T. Jaynes [**46**], for whose quotation we are indebted to Ref.[**47**]:

"*Our present quantum mechanical formalism is not purely epistemological; it is a peculiar mixture describing in part realities of Nature, in part incomplete human information about Nature - all scrambled up by Heisenberg and Bohr into an omelette that nobody has seen how to unscramble. Yet we think that this unscrambling is a pre-requisite for any further advance in basic physical theory*".

Clearly enough, Bohm's probability flux-lines, giving a "visual" representation of the standard solutions of the time-dependent Schrödinger equation, correspond to an average (a "scrambling") taken over the dynamic mono-energetic trajectories proposed in the present work. Our *exact* dynamic trajectories and motion laws may therefore be viewed as the response to Jaynes' "unscrambling pre-requisite" for a *further advance in basic physical theory*.

We conclude that our quantum-dynamical Hamiltonian system (22) allows both to find the "missing link" between *classical* particle dynamics and Bohm's theory (a link whose absence could justify both Einstein's attitude and Cushing's historical contingency) and to satisfy Jaynes' "unscrambling pre-requisite", thus allowing a novel, *non-probabilistic* interpretation of Wave Mechanics.